\documentclass[12pt]{article}
\usepackage{amssymb, amsmath}
\usepackage{latexsym}
\usepackage{psfrag}
\usepackage{dsfont}
\usepackage{graphicx}
\usepackage{rotating}
\usepackage{longtable}
\usepackage{indentfirst} 
\usepackage[margin=1.5cm]{caption}
\usepackage{amssymb, amsmath}
\usepackage{multicol}
\usepackage{makeidx}
\usepackage{float}
\usepackage{array}
\usepackage{boxedminipage}
\usepackage{hyphenat}	
\usepackage{mathrsfs,amsmath} 
\usepackage{graphicx}  
\usepackage{dcolumn}   
\usepackage{bm}        
\usepackage{amssymb}   
\usepackage{color}	   
\usepackage{amsmath}   
\usepackage{algorithm}
\usepackage{algorithmic} 
\usepackage[latin1]{inputenc}
\usepackage{amsfonts}
\usepackage{amssymb}
\usepackage{verbatim}
\usepackage{array}
\usepackage{multirow}
\usepackage{dcolumn}
\usepackage{color}
\usepackage[noadjust]{cite}
\usepackage{url}
\usepackage{graphicx}
\usepackage{subfig}
\usepackage{bbm}
\usepackage{balance}

\usepackage{tabularx,booktabs}

\title{\textbf{Adaptive Filter for Automatic Identification of Multiple Faults in a Noisy OTDR Profile}} 
\author{Jean~Pierre~von~der~Weid,
		Mario~H.~Souto, 
		Joaquim~D.~Garcia,\\
        and~Gustavo~C.~Amaral}

\begin{document}
\maketitle

\begin{abstract}
We present a novel methodology able to distinguish meaningful level shifts from typical signal fluctuations. A two-stage regularization filtering can accurately identify the location of the significant level-shifts with an efficient parameter-free algorithm. The developed methodology demands low computational effort and can easily be embedded in a dedicated processing unit. Our case studies compare the new methodology with current available ones and show that it is the most adequate technique for fast detection of multiple unknown level-shifts in a noisy OTDR profile. 
\end{abstract}

\section{Introduction}\label{Introduction}

The central problem in fiber monitoring is the detection of small faults or losses most commonly performed by inspecting the trace of an Optical Time Domain Reflectometer (OTDR) \cite{BarnoskiAO1977}. These faults appear as small level shifts in a slowly varying backscattered optical power, eventually masked by the detector noise. Averaging over many OTDR shots is usually required to get access to the information needed. However, measurement time is of paramount importance in network monitoring, so that signal processing and filtering is a fundamental tool to improve time and sensitivity of the overall process. Moreover, in the case of wavelength multiplexed optical networks (WDM-PON) the problem is still worse because coherent backscattered power fluctuations (CRN) cannot be averaged out by summing up many OTDR shots \cite{AmaralJLT2015}.

The problem of identifying level shifts may be referred in the literature as \emph{identification of structural breaks}, \emph{step filtering}, \emph{jump detection} or \emph{regime switching} \cite{tsay1988outliers}. One possible approach to identifying such steps would be to decompose the signal into its level and additional components \cite{kalman1960new,durbin2012time,hodrick1997postwar}. However, none of these methods can be directly applied to detecting level-shifts when the presence and position of the shift is unknown. Boyd and Kim \cite{kim2009ell_1} have introduced the $\ell_1$ Trend Filter, a piece-wise linear filter capable of identifying location and magnitude of peaks and shifts. Recently, this technique has been successfully employed in automatically detecting faults in an optical fiber link \cite{AmaralJLT2015}.

In this paper, we analyze the potential of different techniques in identifying level shifts associated to fiber faults in a noisy OTDR profile and we make use of the Tunable OTDR reported in \cite{AmaralJLT2015} for the field tests. The main contributions of this paper are twofold. First, a three-steps extension to Boyd's $\ell_1$ Trend Filter is proposed. The filter's accuracy on detecting the correct level-shift position is improved via a supplementary reweighted $\ell_1$ minimization on the first two steps whereas the bias on the level shift estimate is corrected on the third. Since regularization leads to a shrunk estimator, a traditional least squares estimation of the selected level-shifts can produce unbiased magnitude estimators making the third step imperative. This novel three-steps filter is hereby referred to as the \textit{$\ell_1$ Adaptive Filter}.

The second contribution consists of an efficient algorithm for the first two steps of the $\ell_1$ Adaptive Filter. Each of those steps is a convex optimization problem and there are multiple off-the-shelf techniques for solving it. Nevertheless, the special structure of the problem induces a tailored algorithm. Inspired by the Covariance Update Coordinate Descent \cite{friedman2010regularization}, we developed a faster filtering algorithm designed for this particular problem.

This paper is organized as follows. We present the test bench optical fiber link used throughout the article and its profile acquired by the Tunable OTDR in Section \ref{FiberSection}. Section \ref{textoFILTER} starts by describing the level-shift filter. It is shown how the filtered signal can be obtained by solving a special minimization problem that leads to the Least Absolute Shrinkage and Selection Operator (LASSO) \cite{tibshirani1996regression}, which is the starting point of our tailored algorithm. Section \ref{adaLASSOsection} introduces an adaptive and improved version of the filter based on a reweighted $\ell$-1 minimization and proposes an efficient algorithm for solving the adaptive filter. The comparison between several signal processing techniques and our own is exposed in Section \ref{MethodCompare} where the focus is on the ability of fast and accurate detection of level shifts in the test bench profile. Finally, in Section \ref{Conclusion}, conclusions are drawn over the effectiveness of our novel methodology and some future researches are suggested.

\section{Optical fiber fault detection}\label{FiberSection}
The Optical Time Domains Reflectomettry (OTDR) technique consists of sending a light pulse into an optical fiber and measuring the Rayleigh backscattered light \cite{BarnoskiAO1977}. The OTDR trace is the backscattered power as a function of the distance: a descending line in logarithmic scale with angular coefficient equal to the fiber's attenuation coefficient \cite{agrawal2002book}. Any event which causes optical power loss is interpreted accordingly so fiber faults and defective fiber splices can be identified by abrupt level shifts in the signal profile. Due to its mathematical characteristics, the signal can, therefore, be decomposed into piecewise linear functions with slopes that correspond to the attenuation coefficient of the respective fiber stretch. Recently, a Tunable OTDR has been proposed in which Boyd's $\ell_1$ Trend Filter is used to perform such decomposition and the positions and magnitudes o the faults are automatically identified.

\begin{figure}[H]
\center
\includegraphics[width=0.95\textwidth]{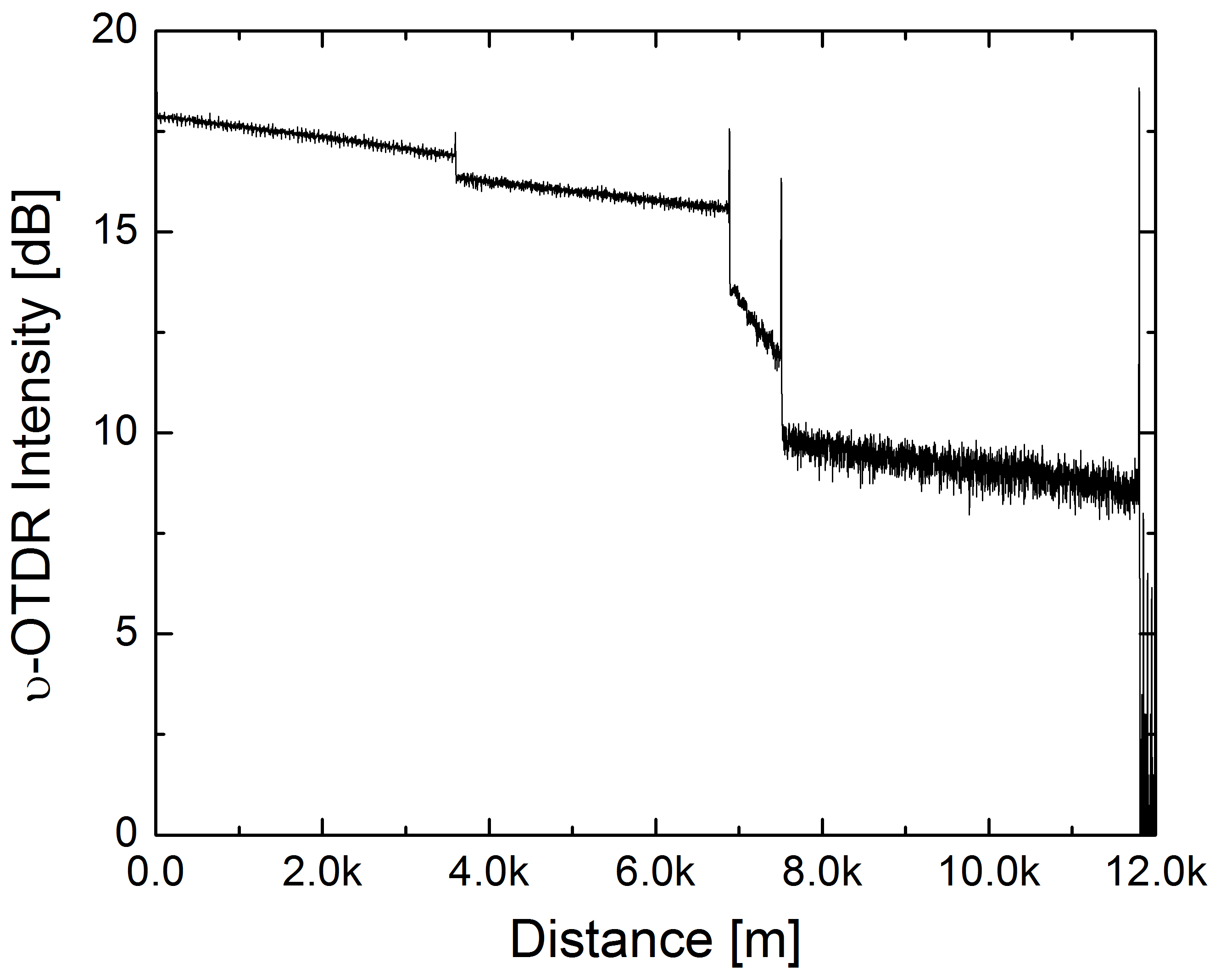}		
\caption{Raw $\nu$-OTDR profile of the testbench $12$ kilometers fiber. Several faults spaced by no more than $100$ meters were induced between $6$ and $8$ kilometers. The first fiber stretch up to $\sim3.5$ kilometers has a different attenuation coefficient which is translated by a linear region with different slope.} 
\label{adaLASSO_Example}
\end{figure}

The achieved spatial resolution reported in \cite{AmaralJLT2015} is $5.72$ meters with a spatial indexation of the data series of $1$ meter due to the hardware maximum clock speed limitation. We employed this technique to measure the profile of a test bench optical fiber link with several fault events specifically designed as to mimic the difficulty in monitoring an unknown fiber link. The testbench fiber has approximately $12$ kilometers and was tailored to include interesting features such as the presence of small faults next to big ones, several faults spaced by no more than $100$ meters, reflective and non-reflective fault events, and fibers with different attenuation coefficients. The OTDR profile is depicted in Fig. \ref{adaLASSO_Example}. 

Two main sources of noise are present in Tunable-OTDR measures: the Poisson Noise associated with the counting process; and the Coherent Rayleigh Noise (CRN) associated with the random interference from backscattered wave fronts. Since the first grows with the square root of the number of detections, the SNR due to Poisson Noise can be set at arbitrary levels by extending the monitoring period. This approach, however, would lead to monitoring periods above the expected from an OTDR technique. The CRN, on the other hand, is an intrinsic effect which cannot be fully eliminated since it is dependent on the probe pulse's linewidth. As commented in \cite{VillafaniIMOC2013}, wavelength sweep techniques can be employed to minimize its contribution but not to eliminate it. In this context, a filtering method which deals with such noise sources mathematically without compromising the technique's performance is highly desirable.

\section{Filtering via the $\ell_1$ norm}\label{textoFILTER}

As introduced in \cite{kim2009ell_1}, the main idea of the $\ell_1$ Trend Filter is to decompose the signal into a piecewise linear curve. To do so, the filter aims to minimize the sum of squared residuals simultaneously penalizing the first difference in the filtered signal to obtain a sparse solution; in this case, sparsely distributed level-shifts.

The most intuitive idea is to directly minimize the number of level-shifts. Unfortunately, this is a combinatorial task and belongs to the class of NP-hard problems \cite{natarajan1995sparse}. With the development of theoretical results \cite{donoho1989uncertainty} in the past twenty years, the idea of using the $\ell_1$ norm in place of the counting norm has gained much attention and acceptance \cite{park2000l1minimization} \cite{oberhofer1982consistency} . Consider a vector $y \in \mathbb{R}^N$ containing the signal observations and let $N$ be the total number of observations. We can define the filtered signal as the combination of a piecewise-linear component and occasional level-shifts. The filtering process can be achieved by solving the following multi-objective optimization problem:

\begin{equation}
\mathop{\min_{(a, \{\theta_k\}_{k=1}^{N})}} \tfrac{1}{2} \sum_{k=1}^{N} (y_k-\theta_k-ak)^{2} + \lambda \sum_{k=2}^{N} |\theta_k-\theta_{k-1}|
\label{L1}
\end{equation}

In this sense, the signal $y$ is decomposed into the sum of a monotonic linear function with the slope $a \in \mathbb{R}$ and the level $\theta \in \mathbb{R}^N$. The penalty term $\lambda \sum_{k=2}^{N} |\theta_k-\theta_{k-1}|$, penalizes shifts on the level component. Thus, the regularization encourages the level to be as constant as possible. This filter can be seen as a particular form of the fused lasso \cite{tibshirani2005sparsity}.

The tuning parameter $\lambda$ is a non-negative real number that controls the frequency and the amount of components used by the filter to describe the signal of interest. By changing the value of $\lambda$ it is possible to manage the trade-off between the total sum of residuals and the number of components of the filtered signal. The optimization problem (\ref{L1}) can be reduced to the problem of finding a sparse solution for a particular minimization problem. Let us consider the matrix $X \in \mathbb{R}^{N \times N}$  which includes all the possible candidates for a fault position in its columns, i.e., the $j^{th}$ candidate is a \textit{step} function of the form $u_{-1}\left(k-j\right)$ which is described, in matrix form, as a vector of ones up to the $j^{th}$ position and zeros in the remaining positions. The simplest form of the matrix $X$ which considers a single slope is presented in Eq. \ref{eq:matrixX}.
\begin{equation}
X = \begin{bmatrix}
 1         & 0         & 0      & \cdots &  0        & 0 \\
 2         & 1         & 0      & \cdots &  0        & 0 \\
 3         & 1         & 1      & \cdots &  0        & 0 \\
 \vdots & \vdots &         & \ddots & \vdots & \vdots \\
 N-1     & 1         & 1      & \cdots & 1         & 0 \\
 N         & 1         & 1      &  \cdots& 1         & 1
\end{bmatrix}\label{eq:matrixX}
\end{equation}

When written as $\mu = X\beta$, the filtered signal $\mu$ is interpreted as the linear combination of selected level shifts which best approximate the original signal $y$. The vector of unknowns $\beta \in \mathbb{R}^N$ corresponds to the component selector: whenever the $i^{\text{\tiny th}}$ position of $\beta$ is different from zero, a level shift component is introduced at position $i$ in the linear combination to form $\mu$. Consequently, the signal $y$ can be stated as $y = X\beta + \varepsilon$, where $\varepsilon$ is the approximation error. With the aforementioned re-parametrization, it becomes clear that the filter (\ref{L1}) is equivalent to a regularized least squares of the form:
\begin{equation}
  \mathop{\min_{(\beta_i)}} \tfrac{1}{2} \| X\beta - y \|^2_2 + \lambda \sum_{i=2}^{N}|\beta_i | \label{L1_lasso}
\end{equation}

The expression in Eq. (\ref{L1_lasso}) is a method of major interest in several fields of study. The most renowned is the LASSO \cite{tibshirani1996regression}, which is used to perform model fitting and variable selection in the high-dimensional statistics framework. Another well-known method is the \textit{Basis Pursuit Denoising} \cite{chen1998atomic}, its equivalent in the signal processing literature, broadly used in sparse signal recovery.

\section{$\ell_1$ Adaptive Filter}\label{adaLASSOsection}

The main idea of the $\ell_1$ Adaptive Filter is to firstly elect potential candidates by finding the vector $\beta_j^{I}$ which minimizes Eq. (\ref{L1_lasso}). Afterwards, each of those elected variables are differently penalized by some parameter that is proportional to $\frac{1}{|\beta_j^{I}|^{\gamma}}$ for some $\gamma >0$. Finally, the selected variables, or components, are given by 

 \begin{equation}
  \mathop{\min_{\beta}} \tfrac{1}{2} \| X\beta - y \|^2_2 + \lambda \sum_{2}^{N}\frac{1}{|\beta_i^{I}|^{\gamma}}|\beta_i |
\label{adaptive_filter}
 \end{equation}
Since the selected components are comprised within the ones elected by Eq. (\ref{L1_lasso}), fewer components are set to be different from zero by the $\ell_1$ Adaptive Filter.

In order to select the best estimators, a two dimensional grid, usually built on logarithmic scale, is created for the pair $\{\rho,\gamma\}$. The best pair needs to be chosen by some metric evaluated at every point of the two dimensional grid. According to Zou \cite{zou2007degrees}, the \textit{Bayesian Information Criterion} (BIC) \cite{schwarz1978estimating} can consistently select the values of the regularizer that will most likely recover the correct set of variables \cite{zhang2010regularization}.

Zou proposed the \textit{Adaptive} LASSO \cite{zou2006adaptive} as an improvement on the ability of identifying significant variables for the problem described by Eq. \ref{adaptive_filter}. The greatest benefit is the technique's consistency in variable selection since the underlying true model should be identified when the number of observations grows. Nevertheless, the LASSO produces biased estimators due to the shrinking process behind the technique \cite{buhlmann2011statistics}\cite{hastie2009elements}. As a result, the filter's capability to perform the signal recovery is diminished. More precisely, the magnitude of every level-shift will be smaller than expected. This last issue can be easily solved by calculating the Ordinary Least Square (OLS) estimation after the adaptative filter has chosen the proper components to describe the signal. Since just a few components will be nonzero, obtaining the OLS estimator will be straightforward and computationally cheap. 

A series of algorithms are known to solve problems of the form presented in Eqs. (\ref{L1_lasso}) or (\ref{adaptive_filter}). Since the two problems are very similar apart from the weighting factor in the latter, they can be solved by the same technique. It is also important to highlight that both optimization problems are convex and therefore any local minimum must be a global minimum \cite{boyd2009convex}. We propose a tailored version of the \textit{Covariance Updates Coordinate Descent} \cite{friedman2010regularization}\cite{friedman2007pathwise} for solving the optimization problem.

The Coordinate Descent algorithm is very popular in the field of high-dimensional statistics due to its excellent computational times. The core of the method is to optimize a multivariate function by sequentially minimizing the objective function with respect to a single coordinate direction instead of the multivariate direction given by the gradient \cite{tseng2009coordinate}. Since the objective function is composed of a differentiable and convex component plus a non-differentiable but separable component, optimality is guaranteed \cite{tseng1988coordinate}. The optimization process works, in every iteration, as to minimize the objective function at a coordinate without changing the values of the remaining ones. The algorithm will perform the core cycle of Algorithm \ref{CoordDesc} until no coordinate has changed its value which means convergence was reached.

\begin{algorithm}[H]
\caption{Coordinate Descent}
\label{CoordDesc} 
\begin{algorithmic}
\STATE{Initialize active set $\mathcal{A} = \varnothing$}
\FORALL{$\lambda_k \in \{\lambda_1,\cdots,\lambda_{max}\}$}
\WHILE{not converge}
\FORALL{$j \in \{1,.., N-1\}$}
\STATE{$\hat{\beta}_j^{ols} =\tfrac{1}{N} \left[ \langle X_j , y \rangle - \displaystyle\sum_{l \in \mathcal{A}}\langle X_j , X_l \rangle \tilde{\beta}_l \right] + \tilde{\beta}_j$}
\IF{$\alpha_{k,i,j} \geq |\hat{\beta}_j^{ols}|$}
\STATE{$\tilde{\beta}_j = 0$}
\STATE{$\mathcal{A} = \mathcal{A} \backslash \{ j\}$}
\ELSE
\STATE{$\tilde{\beta}_j = sign(\hat{\beta}^{ols}_j)(|\hat{\beta}^{ols}_j| - \alpha_{k,i,j})$}
\IF{$j \notin \mathcal{A}$}
\STATE{$\mathcal{A} = \mathcal{A} \cup \{ j\}$}
\ENDIF
\ENDIF
\ENDFOR
\ENDWHILE
\STATE{$\hat{\beta}^{CD}_j(\lambda_k) = \tilde{\beta}_j \hspace{0.5cm} \forall \hspace{0.2cm} j = 0,.., N-1$}
\ENDFOR
\end{algorithmic}
\end{algorithm} 

The Coordinate Descent method presented above can take different forms, namely \textit{Type 1} and \textit{Type 2}, depending on the expression assumed by $\alpha_{k,i,j}$. \textit{Type 1} is characterized by $\alpha_{k,i,j} = \lambda_k  \sigma_j$ whereas, for \textit{Type 2}, $\alpha_{k,i,j} = \frac{\lambda_k  \sigma_j}{|\beta^{I}_j|^{\gamma_i}}$. This is the same as assuming $\beta^{I}_j=1$ for the \textit{Type 1} case. As discussed extensively in \cite{friedman2010regularization}, the computational effort of each cycle depends on the entry of a new component $X_j$ in the model which is accompanied by the computation of the inner products $\langle X_j , X_l \rangle$ ($O\left(N\right)$ operations) and the update of the current coefficients in the model ($O\left(p\right)$ operations) where $p$ is the number of components in the model. Hence, a model that can be described by $m$ components will take $O\left(Nm\right)$ operations. 

Taking advantage of the special structure of the matrix $X$, the inner products necessary for the run of Algorithm \ref{CoordDesc} can be analytically determined as a function of the values of $X$. We dedicate Appendix I to the mathematical development of such analytic expressions and other features as the calculation of the inner products between the columns of the matrix $X$ and the vector $y$. This procedure, also necessary for the run of Algorithm \ref{CoordDesc}, can be executed in a simplified and more efficient manner whenever the implementation has enough memory space available. The storage of the computed inner products in a Gram Matrix, on the other hand, does not to limit the effort associated to the procedure therefore the algorithm can be ran with equal efficiency in a device with limited available memory such as a Micro-Controller or an FPGA.

Algorithm \ref{CoordDesc} is used to obtain the filter's estimate for a given tuning parameter $\lambda$ where the potential values for $\lambda$ are within the range $[0, \lambda_{MAX}]$. The fact that $\lambda_{MAX}$ can be calculated in beforehand renders the algorithm completely parameter-free. Let $\sigma$ be the vector containing the standard deviation of each column of matrix $X$. Then, 
\begin{align}
\begin{aligned}
\lambda_{MAX} = \left\{ \lambda = \bigg| \sum_{i=1}^N\right. &X_{ij} y_i \bigg| : \\
\lambda &\left.\geq \bigg| \sum_{i=1}^N \frac{\sigma_k}{\sigma_j} X_{ik} y_i \bigg| \hspace{0.1cm} \forall \hspace{0.1cm} k \neq j \right\}
\end{aligned}
\label{sigma_max}
\end{align}
When $\lambda = 0$, the solution will be the same as the OLS estimate with a great amount of level-shifts. On the other hand, when $\lambda = \lambda_{MAX}$, there will be no components to describe the signal. These are the two extreme cases of the LASSO grid. It is important to note that the calculations of $\lambda_{MAX}$ should be performed before the normalization discussed in Appendix I. The final solution algorithm for the $\ell_1$ Adaptive Filter is presented in Algorithm \ref{adaLasso} for which Algorithm \ref{CoordDesc} is the core.

\begin{algorithm}[H]
\caption{$\ell_1$ Adaptive Filter}
\label{adaLasso} 
\begin{algorithmic}
\STATE{Initialize $\tilde{\beta}_j = 0 \hspace{0.5cm} \forall \hspace{0.2cm}  j = 0,.., N-1$}

\STATE{}
\STATE{\textbf{Run Algorithm 1}\text{ Type 1}.}
\STATE{Obtain: $\hat{\beta}_j^{CD}(\lambda_k) \hspace{0.5cm} \forall \hspace{0.2cm} j = 0,.., N-1 \textit{ and } k=1,.., m$.}
\STATE{Define: $\hat{\beta}^{I}$ as the best $\beta^{CD}(\lambda_k)$ }

\STATE{}
\FORALL{$\gamma_i \in \{\gamma_1,\cdots,\gamma_q\}$}

\STATE{\textbf{Run Algorithm 1}\text{ Type 2}.}
\STATE{Obtain: $\hat{\beta}^{ADA}_j(\lambda_k,\gamma_i) =\hat{\beta}_j^{CD}(\lambda_k)$}
\STATE{$\hspace{2.9cm} \forall \hspace{0.2cm} j = 0,.., N-1 \textit{ and } k = 1,.., m$.}

\ENDFOR

\STATE{Define: $\hat{\beta}$ as the best $\hat{\beta}^{ADA}(\lambda_k,\gamma_i)$}
\STATE{Compute: OLS estimator for $\hat{\beta}^{ADA}(\lambda_k,\gamma_i) \neq 0$}
\end{algorithmic}
\end{algorithm} 

For simplicity, the algorithm is initialized with $\hat{\beta}^{CD}_j(\lambda_k) =0$, $\forall \lambda_k \in \{\lambda_1,\cdots,\lambda_m\}$. This enhances the efficiency of the algorithm since it saves many comparisons and memory accesses. The interpretation for this event is that the $j^{th}$ value of $\beta_j^{I}$ was chosen to be irrelevant during initialization.

\section{Results and Discussion}

\begin{figure}[H]
\center
\includegraphics[width=0.95\textwidth]{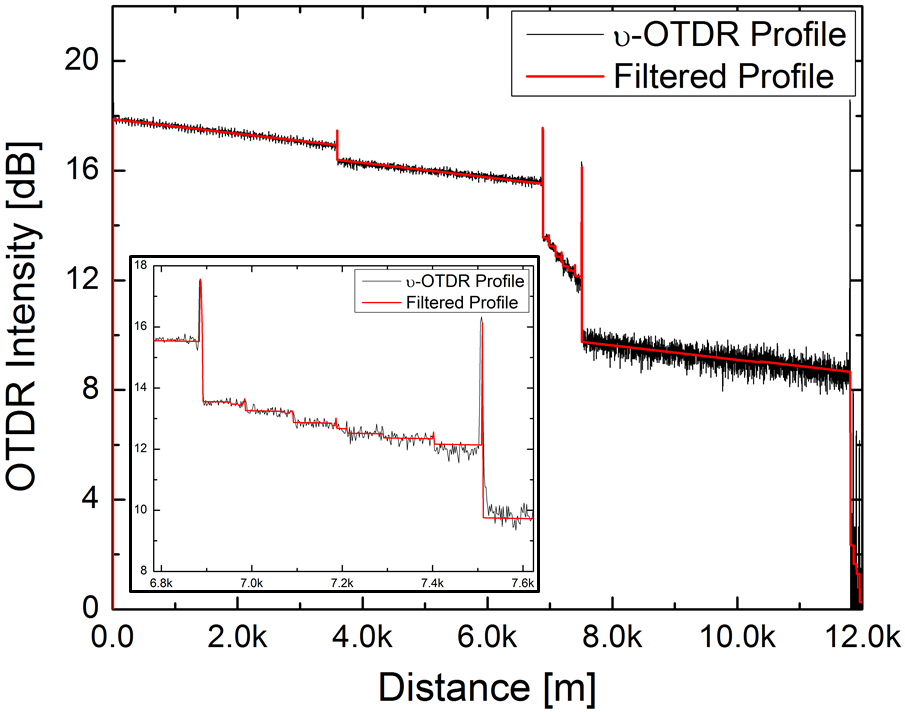}		
\caption{Raw $\nu$-OTDR Profile (black) and $\ell_1$ Adaptive Filter result (red). The inset details five consecutive faults ranging from $0.1$ to $0.3$ dB within a $600$ meters stretch which were accurately detected by the algorithm.} 
\label{adaL1_Result}	
\end{figure}

In Fig. \ref{adaL1_Result}, we show the filtered OTDR profile when the $\ell_1$ Adaptive Filter is employed. The algorithm is capable of accurately selecting the induced faults, even those as small as 0.1 dB. Another feature of the algorithm is that faults close to each other are unequivocally detected as detailed by the inset of Fig. \ref{adaL1_Result}.  This result is important to attest the sensibility of the method even in the presence of consecutive small faults next to big faults.

\subsection{Comparison between different Methodologies}\label{MethodCompare}

From the result of Fig. \ref{adaL1_Result}, it becomes clear that the $\ell_1$ Adaptive Filter represents a potential candidate for optical fiber fault detection applications. However, whether or not it is the best candidate depends on the performance of others techniques aimed at the same purpose. Therefore, we study the $\ell_1$ Adaptive Filter capacity on detecting fiber faults against recent mathematical approaches such as the Potts Functional \cite{storath2014jump}, and classical ones such as the ART \cite{rea2010identification} and the $\ell_1$-style minimization \cite{kim2009ell_1,candes2008enhancing} which was the technique presented in \cite{AmaralJLT2015} in conjunction with the Tunable OTDR for automatic fault detection. A total of six different methodologies were compared: the L1 Potts functional; the L2 Potts Functional; the reweighted L1 minimization proposed in \cite{candes2008enhancing}; the ART using BIC as a pruning criterion; the LASSO; and the proposed $\ell_1$ Adaptive Filter. 

The L1 Potts, the L2 Potts, the reweighted L1 and the LASSO depend on a tuning parameter so a grid of size $100$ was created and the best model was chosen via the BIC. The $\ell_1$ Adaptive Filter uses the LASSO estimator as an initial condition therefore it is subject to the LASSO grid. A simpler version of the problem, with no linear trend, was used as input for both the L1 and L2 Potts algorithms since they do not account for slopes. Comparisons between the methodologies were drawn in the light of four main parameters: elapsed time; number of unsuccessful detections (within the resolution of the technique); and number of spurious detections. These were elected given the ultimate goal of a fiber monitoring method: fast and accurate detection of a fault.

The $\nu$-OTDR acquisition period is intimately tied to the error associated to the number of count events at each position due to the Poisson Noise as commented in Section \ref{FiberSection}. A reasonable SNR is therefore necessary so the competing algorithms can truly be tested in its prowess to identify fiber characteristics in an OTDR profile. The testbench fiber presented in Figs. \ref{adaLASSO_Example} and \ref{adaL1_Result} demands a minimum of $100$ seconds acquisition period in order for the last portion of the fiber link to be well above the noise floor in the OTDR profile which correspond to approximately $8 \cdot 10^5$ launched probing pulses. The rate at which the monitoring technique launches the probe pulses depend on the fiber's length due to a restriction of 1 pulse at a time traversing the fiber \cite{AmaralJLT2015}. In the presented case, the rate is $8$ kHz for the $\sim12$ kilometer-long testbench fiber.

\begin{figure}[ht]
\center
\includegraphics[width=0.95\textwidth]{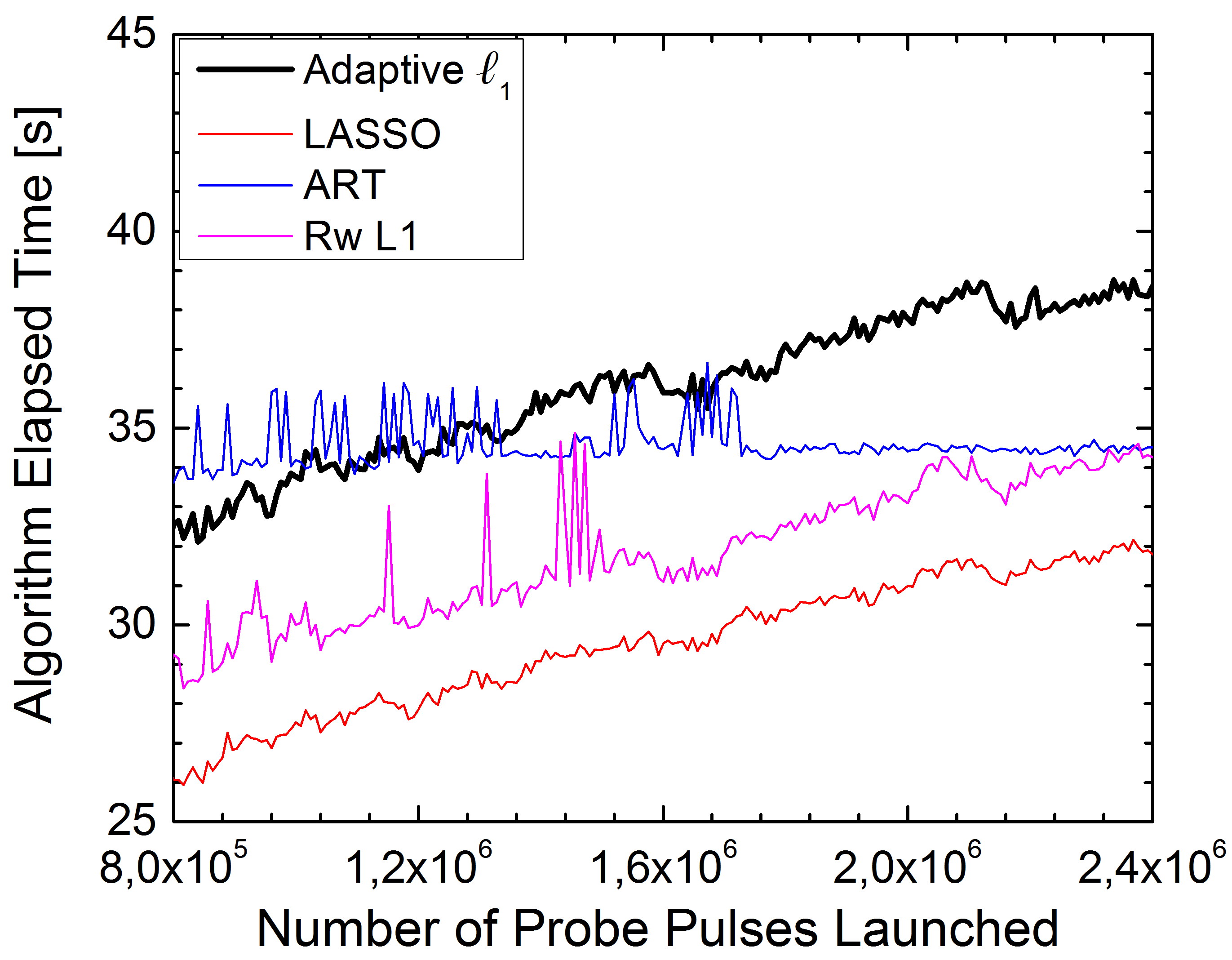}
\caption{Algorithm elapsed time versus the number of probing pulses launched into the testbench fiber. Probe pulse rate is $8$ kHz. The L1 and L2 Potts Functional yield higher than one minute processing time so their results were subtracted from the graph to ease the visualization.} 
\label{fig:adaL1_Result1}
\end{figure}

\begin{figure}[ht]
\center
\includegraphics[width=0.95\textwidth]{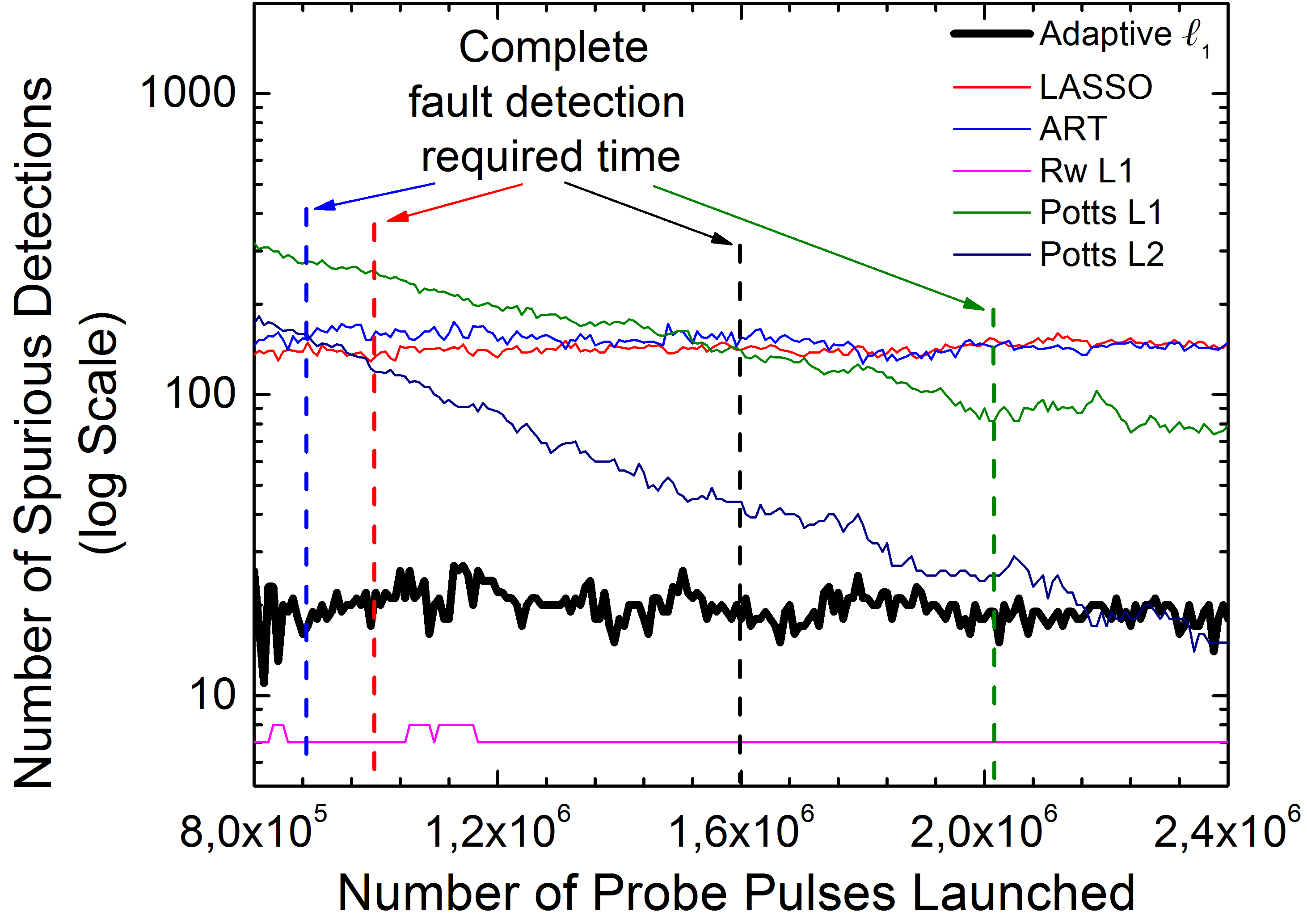}
\caption{Number of spurious detections versus the number of probing pulses launched into the testbench fiber. Probe pulse rate is $8$ kHz. This feature represents the accuracy of the filter in detecting the faults of the testbench fiber. The traced lines correspond to the minimum time required by each technique for detecting all faults present in the testbench link. The Reweighted L1 and Potts L2 algorithms were not able to identify all faults within $300$ seconds of acquisition time.}
\label{fig:adaL1_Result2}
\end{figure}

The acquired data sets from $100$ seconds up to $300$ seconds ($8\cdot10^5$ to $2.4\cdot10^6$ launched probe pulses respectively) -- with 1 second spacing between each sample -- were fed into the listed algorithms so their performance could be assessed. The graphs depicted in Figs. \ref{fig:adaL1_Result1} and \ref{fig:adaL1_Result2} trace the algorithms elapsed time and the number of spurious detections versus number of probing pulses launched into the testbench fiber. Due to the difficulty in graphical visualization, the number of missed faults is not displayed in a separate graph, but rather along with the results of the spurious detections in Fig. \ref{fig:adaL1_Result2}.

A few comments should be included before the results are discussed. The Reweighted L1 is unable to detect all the consecutive faults between $6$ and $8$ kilometers. The Potts L2, despite being able to detect some of these faults, fails to detect all of them. These results are independent on the acquisition time. Both the LASSO and the ART algorithms detect all the induced faults with minimum acquisition time. The Potts L1 and $\ell_1$ Adaptive Filter get more sensitive as the acquisition time rises but the first takes as long as $\sim250$ seconds to detect all of the faults whereas the second reaches this mark with $\sim200$ seconds.

Analysis of Fig. \ref{fig:adaL1_Result1} indicates that the $\ell_1$ Adaptive Filter shows comparable timing characteristics to the fastest methods with no more than $10$ seconds difference. The increase in elapsed time as the acquisition time rises, even though counter intuitive at first, can be attributed to the convergence criterion of the \emph{Coordinate Descent}: as the faults become more distinctive, the selection between two candidates separate by few meters gets harder due to numerical issues. There are methods which deal with such numerical impairments and can be employed to accelerate the algorithm's convergence \cite{}. The Potts L1 and L2 algorithms exhibit higher than one minute elapsed times and its curves were, therefore, subtracted from the graph to ease the visualization. 

Fig. \ref{fig:adaL1_Result2} conveys a good deal of information regarding the accuracy of each method. Those methodologies which achieved the lowest number of missed faults, i.e., required the least acquisition time to identify all the faults in the link, namely the LASSO and ART algorithms, are also the ones with the highest number of spurious detections as depicted by the \textit{red} and \textit{blue} lines of Fig. \ref{fig:adaL1_Result2}. Conversely, the Reweighted L1 algorithm shows the smallest number of spurious detections but is unable to detect the five consecutive fault events which are detailed in the inset of Fig. \ref{adaL1_Result}. 

Data acquisition, as commented in Section \ref{FiberSection}, is performed in $1$ meter bins even though the spatial resolution of the Tunable OTDR technique is $5.72$ meters. For that reason, a fault event is scarcely characterized by a single step within a $1$ meter slot but rather represented by a cluster of steps inside a $6$ meter range. In order to simplify our comparison method, all fault positions identified by the algorithms which do not match the value indicated by the standard OTDR profile serving as reference, are considered spurious. Therefore, the apparent high number of spurious detections of the $\ell_1$ Adaptive Filter is actually an expected result. Considering an average of 3 spurious detections per cluster (smaller faults tend to be represented by less than a $6$ meter clusters), we find that the $\ell_1$ Adaptive Filter hardly identifies a step outside a cluster and figures as the most accurate technique among those presented. The association of level shift clusters with a fault has already been dealt with in \cite{AmaralJLT2015}.

For the presented testbench, the $\ell_1$ Adaptive Filter was able to include all faults in the optical link from $200$ seconds acquisition time with the benefit of a minimum number of spurious detections. The algorithm's precision, accuracy, competitive timing (total monitoring time including acquisition and processing steps is $\sim 245$ seconds), and readily applicability in the architecture presented in \cite{AmaralJLT2015} sets the $\ell_1$ Adaptive Filter as a promising candidate for noisy OTDR filtering and fault detection method. The $\ell_1$ Adaptive Filter presents the best compromise between the number of spurious detections and the ability to identify the faults present in the link with reasonable processing time.

\section{Conclusion}\label{Conclusion}

We have shown that the proposed $\ell_1$ Adaptive Filter outperforms well-established methods found in the literature that tackle the problem of detecting level shifts in a data stream. The main contribution of our study is a novel successful tool for detecting multiple level-shifts at unknown times in a signal of interest with direct applications in the filtering and automatic identification of faults in a noisy OTDR profile. 

Additionally, we proposed an algorithm based on the Coordinate Descent which offers enhanced selection accuracy and competitive timing. Inspection of a testbench fiber optical link profile with different method shows that the $\ell_1$ Adaptive Filter presents the best compromise between accurate and fast detection of fault events. The low computational effort demanded by the algorithm allows it to be embedded in dedicated processing units such as an FPGA or Micro-Controller.

Among the many possible future research directions we mention two: i) Inclusion of a spike trend filter, along with the step trend filter, in order to discern between a reflective and a non-reflective fiber fault and also to eliminate pitfalls at the final portion of the fiber which contains several spurious detections.; ii) Exploration of the parallel computing capabilities of the Coordinate Descent algorithm to increase the time response of the filter.

\section*{Appendix I}\label{AppendixI}

It is important to observe that the matrix $X$, as defined in Eq. \ref{eq:matrixX}, is not fit to be fed into Algorithm \ref{CoordDesc} since its columns ought to be statistically normalized in order to have comparable candidates of explanatory variables. In order to do so, we have to subtract each column of $X$ by its mean and divide each of them by their own standard deviation. This normalization has an analytic form as follows.

We label the lines and columns of the $N \times N$ matrix $X$ with the numbers $0$ to $N-1$. Both mean and standard deviation of the $i^{th}$ column are respectively given by:

\begin{displaymath}
\mu_i = \left\{ \begin{array}{ll}
\frac{N+1}{2}  \hspace{0.6cm}\textrm{if i = 0,}\\
\\
\frac{N-i}{N}  \hspace{0.6cm}\textrm{otherwise.}
\end{array}\right.
\end{displaymath}
\begin{displaymath}
\sigma_i = \left\{ \begin{array}{ll}
\sqrt{\frac{1}{6} (N+1)(2N+1)-\mu_0^2}  \hspace{1.05cm}\textrm{if i = 0,}\\
\\
\sqrt{\frac{1}{N}  ( i \mu_i^2 +(N-i)(1-\mu_i)^2 )}  \hspace{0.6cm}\textrm{otherwise.}
\end{array}\right.
\end{displaymath}
Then, we define $U_i=\frac{-\mu_i}{\sigma_i}$ and $L_i=\frac{1-\mu_i}{\sigma_i}$, for $i\neq0$ with the following interpretation: $U_i$ and $L_i$ are the values of the positions occupied by zeroes and ones on the column $i$ of $X$, respectively, after the above mentioned normalization. Since $U$ is an upper diagonal and $L$ a lower diagonal matrix, both can be stored simultaneously in a single $N \times N$ matrix. Finally, the terms that comprise the Gram Matrix $GM$, i.e., the matrix containing the inner products between every pair of columns of X, are given by:
 
 \vspace{0.05 in}
 $GM_{00}=N-1$
 \vspace{0.05in}
 
 $GM_{0j}=\frac{1}{\sigma_0} \bigg[ j U_j \big[\frac{j+1}{2} -\mu_0\big] + (N-j)L_j \big[\frac{j+1+N}{2} -\mu_0\big] \bigg]$
 \vspace{0.05in}
 
 $GM_{j0}=GM_{0j}$
 \vspace{0.1in}
 
 $GM_{ij}=min(i,j)U_i U_j +(max(i,j)-min(i,j))L_i U_j +$
 
 \hspace{1.3cm}$(N-max(i,j))L_iL_j$
 \vspace{0.05in}
 
Therefore, depending on the size of the problem and the physical memory available for the algorithm, there are two possibilities: calculate the values with the above formulas as they become necessary, or pre-compute the Gram Matrix and store the results. These closed formulas are much faster than computing all the inner products in standard way, so there is gain in time for both approaches.
  
The run time of Algorithm \ref{CoordDesc} can also be optimized if one pre-computes the values $\langle X_j , y \rangle$, the inner products between each column of the normalized $X$ matrix and the normalized $y$ vector. The inner products $\langle X_j , y \rangle$ can be more efficiently computed by calculating $\langle X_0 , y \rangle$ in standard manner and then using Algorithm 1 for $\langle X_i , y \rangle$.

\begin{algorithm}[H]
\caption{Fast $y$ Inner-Products}
\label{Algorithm 1} 
\begin{algorithmic}
\STATE{Initialize $sumY=0$}
\FOR{$i \textbf{ from }  1 \textbf{ to }   N-1$}
\STATE{$sumY=sumY+y_i$}
\STATE{$\langle X_j , y \rangle=sumY(U_i-L_i)$}
\ENDFOR
\end{algorithmic}
\end{algorithm}

\section*{Acknowledgment}
The authors wish to thank brazilian agencies CPNq, CAPES and FAPERJ for financial support.


\bibliographystyle{IEEEtran}
\bibliography{AdaLASSORefs}

\end{document}